\documentclass{bioinfo}
\usepackage{array}
\usepackage{float}
\copyrightyear{2017}
\pubyear{2017}

\begin{document}
\firstpage{1}

\title[SANA for Biological Network Alignment]{An introductory guide to aligning networks using SANA, the Simulated Annealing Network Aligner}
\author[Wayne B. Hayes]{Wayne B. Hayes\footnote{to whom correspondence should be addressed ({\tt whayes@uci.edu})}\;}
\address{Department of Computer Science, University of California, Irvine CA 92697-3435, USA}

\history{This is a preprint. Citation to published version below.}

\editor{{\bf Citation:} Hayes, Wayne B. ``An Introductory Guide to Aligning Networks Using SANA, the Simulated Annealing Network Aligner.'' In {\it Protein-Protein Interaction Networks}, pp. 263-284. Humana, New York, NY, 2020.
}

\maketitle

\begin{abstract}
%{\bf INCLUDE INDEX, NDEX (they're different) \& OTHER NEW ONES EVEN IF ONLY A TABLE OR ONLY EC S3, and discuss any that we don't include, and why.}

Sequence alignment has had an enormous impact on our understanding of biology, evolution, and disease. The alignment of biological {\em networks} holds similar promise. Biological networks generally model interactions between biomolecules such as proteins, genes, metabolites, or mRNAs. There is strong evidence that the network topology--- the ``structure'' of the network---is correlated with the functions performed, so that network topology can be used to help predict or understand function. However, unlike sequence comparison and alignment---which is an essentially solved problem---network comparison and alignment is an NP-complete problem for which heuristic algorithms must be used.

Here we introduce SANA, the {\it Simulated Annealing Network Aligner}. SANA is one of many algorithms proposed for the arena of biological network alignment. In the context of global network alignment, SANA stands out for its speed, memory efficiency, ease-of-use, and flexibility in the arena of producing alignments between 2 or more networks. SANA produces better alignments in minutes on a laptop than most other algorithms can produce in hours or days of CPU time on large server-class machines.  We walk the user through how to use SANA for several types of biomolecular networks.

\textbf{Availability:} {\tt https://github.com/waynebhayes/sana}

\textbf{Contact:} whayes@uci.edu

\textbf{Supplementary information:} Available online.
\end{abstract}

\section{Introduction}
A biological network consists of a set of nodes representing entities, with edges connecting entities that are related in some way. They come in many varieties, such as protein-protein interaction (PPI) networks \citep{williamson2010protein,jaenicke2012protein}, gene regulatory networks \citep{davidson2010regulatory,karlebach2008modelling}, gene-$\mu$RNA networks \citep{chen2007evolution,prescott2012cell,farazi2013micrornas,kotlyar2015integrated,tokar2017mirdip}, metabolic networks \citep{fiehn2002metabolomics}, brain connectomes \citep{milano2017extensive}, and many others \citep{junker2011analysis}. It is believed that the {\em structure} of the networks, in the form of the network topology, is related to the function of the entities \citep{davidson2010regulatory,DavisPrzulj2015TopoFunction,sporns2010networks}. The {\em alignment} of such networks aims to use connectivity between nodes---the {\em topology} of the network---to aid extraction of information about the nodes and their function. Network alignments can be used to build taxonomic trees and find highly conserved pathways across distant species \citep{GRAAL}; and by extension finding such topological similarities may aid in transfering functional knowledge from better-understood species to less well-understood ones, much like how sequence alignment has been doing so for sequence for decades. Networks are even starting to have an influence on individual human health \citep{van2013whole}

Network alignment is a fundamentally difficult problem: it is a generalization of the  NP-Complete subgraph isomorphism problem \citep{cook1971complexity,GareyJohnson}; and adding to the difficulty is that current data sets are very noisy \citep{falsepositives}. Therefore, modern alignment algorithms try to approximate solutions using heuristic approaches.

There are several sub-classes of network alignment. {\it Global Network Alignment} (GNA) is the task of attempting to completely align entire networks to each other; GNA applied to just two networks is called {\it pairwise} GNA \citep{GRAAL,LGRAAL,MAGNA,MamanoHayesSANA,HubAlign,WAVE,GHOST}, while aligning more than two whole networks is called {\it multiple} GNA. In contrast, {\it Local Network Alignment} (LNA) attempts to find similarity in the local wiring patterns among small groups of nodes, either in the same network, or across many networks.  In all of these cases, alignments can map nodes 1-to-1, or many-to-many; the latter is more biologically realistic since, for example, one gene in yeast may have multiple homologs in mammals. However, the 1-to-1 assumption makes programming simpler and so the majority of aligners take the 1-to-1 mapping as a simplifying assumption.  A more recent version of network alignment looks into modeling dynamic networks (see for example \cite{vijayan2017aligning}).
An excellent comprehensive survey of all these types of alignments is provided by \cite{faisal2015post}.
SANA was originally a 1-to-1 pairwise global network alignment algorithm, although we here also introduce a prototype multiple network alignment version.

\subsection{User/System requirements}
Source code to SANA is available on GitHub at

\noindent{\tt http://github.com/waynebhayes/SANA}, and is best cloned from github on the Unix command line using

\noindent{\tt git clone http://github.com/waynebhayes/SANA}

SANA is written in C++ and runs best on the Unix command line. It has been tested with gcc 4.8, 4.9, 5.2, and 5.4, and runs on Unix, Linux, Mac OS/X, and under the Windows-based Unix emulator Cygwin ({\tt http://cygwin.com}), 32-bit or 64-bit. SANA has a rudimentary Web interface at {\tt http://sana.ics.uci.edu}, and a rudimentary SANA app is available in the Cytoscape app store. SANA expects its input networks to be in a two-column ASCII format we call {\it edge list format}: each line is one edge, specified by listing the two nodes at each end of the edge in arbitrary order (unless {\tt -nodes-have-types} is specified, see below). Duplicate edges and self-loops are not allowed. We also supply a program called {\tt createEdgeList} that can convert the following types of formats into SANA's edgeList format: XML, GML, LEDA, .gw, CSV, LGF.

\subsection{Alignment measures \& objective functions}
\label{sec:measures}

An {\it alignment measure} is any quantity designed to evaluate the quality of a network alignment.  Alignment measures can be classified along many axes.

\subsubsection{Objective vs. non-objective measures.}
The first axis is the distinction between {\em objectives} and what we call {\em post-hoc} measures. While both can be evaluated on any given alignment, any measure used to {\em guide} an alignment {\em as it is being created} is called an {\em objective function}; any measure not used to guide the alignment is generally applied after-the-fact as an independent measure of quality. A good alignment algorithm should be able to use virtually any measure as an objective, and also evaluate the alignment after-the-fact using any other measures which were not used as objectives.

\subsubsection{Graph topology vs. biological measures.} Another axis along which measures can be classified is {\it topological} vs {\it biological}.
\paragraph{A topological measure} quantifies a network alignment based solely on graph-theoretic grounds. Several such measures exist: $EC$ \citep{GRAAL}, $ICS$ \citep{GHOST}, and $S^3$ \citep{MAGNA} quantify the number of edges in one network that are mapped to edges in the other network(s); they are all described in more detail below. Other topological measures use graphlets to quantify local structure \citep{Przulj2004,Tijana2008,Przulj2014HiddenLanguage,LGRAAL}, while still others use graph measures such as spectral analysis \citep{GHOST} and degree similarity-based measures such as Importance \citep{HubAlign}.

\paragraph{Biological measures.}
In contrast, biological measures are usually used to compare the nodes from different networks that have been paired together by the alignment. For genes or proteins, a common measure is the sequence similarity or BLAST score between the aligned nodes \citep{BLAST}; sequence similarity is also frequently combined with topology to produce a hybrid objective function (see for example \cite{MIGRAAL,MAGNA,MamanoHayesSANA,LGRAAL}, among many others). Another biology-based measure is the {\em functional} similarity between pairs of aligned proteins as expressed by GO (Gene Ontology) terms \citep{GO}. While many authors quantify the functional similarity exposed by an alignment using the mean value of various pairwise GO similarity measures across the alignment, such mean-of-pairwise-scores assume each pair of aligned proteins is independent of all others, which is not true in an alignment since every pair is implicitly related to every other pair via the alignment itself. This problem is alleviated by the NetGO score as implemented in SANA \citep{hayes2017sana}, which is a global rather than local scoring mechanism (see below for the meaning of local vs. global measures).

\subsubsection{Local vs. global measures.} The final axis along which network alignment measures can be classified is what we refer as {\em local} vs. {\em global} measures.
\paragraph{A local measure} is one that involves evaluating node pairs that are aligned to each other, and has no explicit dependence on the alignment edges and thus has no explicit dependence on network topology.  Examples of local measures include sequence similarity and pairwise GO term similarity as described above; some local measures such as graphlet similarity \citep{GRAAL,LGRAAL,MAGNA} and Importance \citep{HubAlign} include topology indirectly by pre-computing all-by-all pairwise local topological similarities between all pairs of nodes in one network and all pairs of nodes in the other.

\paragraph{Global measures} are ones that implicitly or explicitly can be computed only on the entire alignment and have nothing to do with pairwise node similarities.  The most common global measures are $EC$, $ICS$, and $S^3$, described in more detail below.

\subsection{Major Topological Measures}
\subsubsection{A useful analogy for topological measures.} 
\begin{figure}[hbt]
\centering
\includegraphics[width=0.99\linewidth]{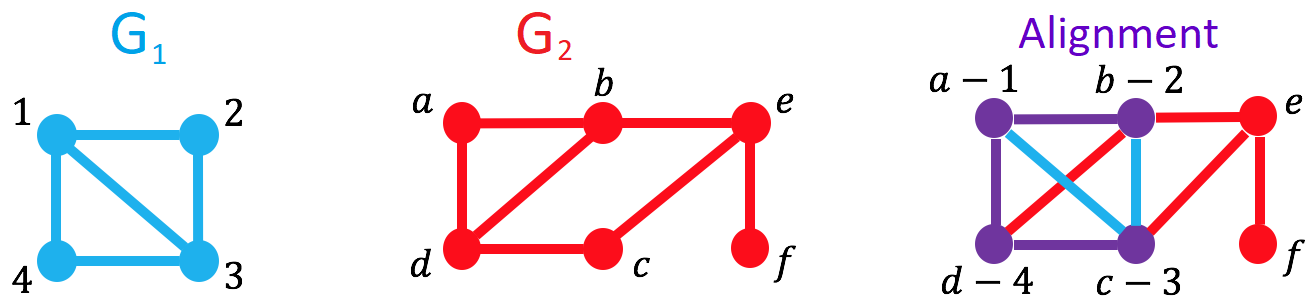}
\caption{A simple example of a network alignment. The smaller network $G_1$ (far left) has its pegs, numbered 1--4, and edges (``laser beams'') depicted in blue; the larger network $G_2$ (middle) has its holes and painted edges depicted in red. One possible alignment (in this case the ``visually obvious'' one) is depicted at the far right.  Here, aligned nodes and edges are depicted as purple; unaligned laser beams from $G_1$ are still blue, and unaligned holes and edges from $G_2$ are still red. As stated in the text, in an alignment figure like the one on the right, the number of edges in $G_1$ is always $m_1=$(purple + blue edges), and the number of edges in $G_2$ is always $m_2=$(purple + red edges). Thus, from the figure, it can be easily seen that $EC=3/5$, and $S^3=3/6$ (where 6 is the total number of edges visible across all colors on the subgraph induced by the alignment); also $ICS=3/4$, since there are 4 edges induced in $G_2$ by the alignment (ie., by purple nodes). The purple network is called the {\it Common Subgraph}, and it can consist of several connected components.  In this case there is only one {\it Common Connected Subgraph} consisting of 4 nodes and 3 edges.
}
\label{fig:purpleAndBlueAndRed}
\end{figure}

In order to more easily understand and discuss topological measures, we introduce an analogy between pairwise network alignment, and the old board game of {\it Battleship}. A Battleship game consists of many holes in a board, and some pegs that are placed into the holes. In our analogy, assume $G_1$ is a ``smaller'' network with $n_1$ nodes and $m_1$ edges, and $G_2$ is a ``larger'' network with $n_2$ nodes and $m_2$ edges, and we assume that $n_1\le n_2$---that is, $G_1$ is the smaller network in terms of number of nodes. We will furthermore depict $G_1$ as blue and $G_2$ as red.
Consider Figure \ref{fig:purpleAndBlueAndRed}: this board has $n_2=6$ red holes with red edges painted between two holes if there is an edge between the two corresponding nodes in $G_2$. The smaller network $G_1$ is represented by $n_1=4$ blue pegs; edges between the pegs are represented by blue ``laser beams'' between the corresponding pegs (because laser beams don't get tangled as pegs are moved from hole to hole). Any placement of the $n_1$ pegs into the $n_2$ holes represents an alignment between $G_1$ and $G_2$; for now we assume that each peg is placed into exactly one hole, so that there are exactly $n_2-n_1$ empty holes. Furthermore, since mixing red and blue creates purple, we depict the alignment (far right of Figure \ref{fig:purpleAndBlueAndRed}) in purple: a blue peg in a red hole is purple, and a blue edge lying on top of a red one is also depicted as purple.

\subsubsection{Edge-based measures: $EC,ICS,S^3$}
We can now define some edge-based topological measures based on this analogy. The fraction of laser beams that lie on top of painted edges is called the $EC$\footnote{Variously called Edge Coverage, Edge Correspondence, or Edge Correctness by various authors} \citep{GRAAL}. The numerator of $EC$ is the number of (purple) edges that are aligned between the two networks, call it $AE$ (an integer), while the denominator is $m_1$; note that since at most $m_1$ edges can be aligned, the value $EC=AE/m_1$ is always less than or equal to 1. The authors of MAGNA \citep{MAGNA} noted that $EC$ is asymmetric: in particular, if $n_1=n_2$ then we can ``turn the board upside down'', swapping the roles of pegs and holes. In that case, the $EC$ changes because $G_1$ and $G_2$ are swapped: in particular, the numerator is always the number of aligned edges $AE$, but the denominator switches from $m_1$ to $m_2$.

The authors of MAGNA fixed the asymmetry of $EC$ by introducing the {\it Symmetric Substructure Score} or $S^3$. Consider the rightmost section of Figure \ref{fig:purpleAndBlueAndRed}, which depicts a proposed alignment. In our analogy, if we ``look down'' on the alignment from above, we can see four different types of edges. There are: (i) $AE$ aligned (purple) edges; (ii) $UE_1$ unaligned (blue) edges from $G_1$; (iii) $UE_{2in}$ unaligned (red) edges in $G_2$ induced between purple nodes; and (iv) $UE_{2out}$ unaligned (red) edges outside the alignment (ie., not induced between purple nodes). Note that the following equations always hold: $m_1=AE+UE_1$ and $m_2=AE+UE_{2in}+UE_{2out}$. Whereas $EC=AE/m_1$, $S^3$ is defined as $AE/(AE+UE_1+UE_{2in})$, and is thus symmetric with respect to the interchange of $G_1$ and $G_2$.
Another way of saying this is that both $EC$ and $S^3$ are rewarded for purple edges in the numerator, but $EC$'s denominator is penalized only for blue edges in its denominator, whereas $S^3$ is penalized in its denominator for both blue and red edges induced by the alignment.

Another measure called ICS {\it Induced Conserved Substructure} \citep{GHOST} measures $AE$ divided by the number of painted edges that exist only between holes that have pegs in them. ICS has the significant disadvantage that it can be maximized by finding a network alignment that {\em minimizes} the number of edges between filled holes\citep{MAGNA,MAGNA++,MamanoHayesSANA}, which can hardly be said to be a good alignment. Consider again Figure \ref{fig:purpleAndBlueAndRed}. The reason ICS is a bad measure is because we could make it equal to $2/2$, ie. 1, by moving node 2 to align with $e$ and 3 to align with $f$; then there would be 2 purple edges ($a$-1 to $d$-4, and $e$-2 to $f$-3) and no red edges {\em induced by the alignment} on $G_2$, even though there would be 3 blue edges (1-2, 4-3, and 1-3) unaligned from $G_1$. Thus there exists an alignment with $ICS=1$ even though it only exposes 2 edges of common topology, which is less common topology discovered by maximizing either $EC$ or $S^3$. This demonstrates the general principle that {\it choosing the right objective function is {\bf crucial} to getting good alignments}.

\subsubsection{Graphlet-based measures.}
Graphlets \citep{Przulj03,Przulj2004} are small, connected, induced subgraphs on a larger graph. They have myriad uses, such as quantifying global topological structure \citep{Przulj2004,Przulj2014HiddenLanguage}. Enumerating graphlets in a large graph is an $NP$-hard problem and much work has gone into heuristics to make their enumeration more efficient. SANA uses ORCA \citep{ORCA} to exhaustively enumerate graphlets in a network. By computing an {\it orbit degree vector} \citep{Tijana2008}, one can create a local measure that compares the orbit degree vectors of two nodes (one from each network); that local measure can then be used as an objective to guide the alignment. GRAAL \citep{GRAAL} was the first to use orbit degree vectors\footnote{In the GRAAL paper we used the term ``graphlet degree vector'' but it's more correctly called an ``orbit degree vector'' because it's a vector of orbit counts, not graphlet counts.}, and SANA uses the exact same mechanism. However, as networks grow larger, the exhaustive enumeration of its graphlets is becoming very expensive. For example, ORCA takes more than 24 hours to compute the orbit degree vectors when aligning the 2018 BioGRID \citep{chatr2017biogrid} networks of {\it H. sapiens} and {\it S. cerevisiae}. Instead, we intend to move SANA towards statistical sampling of graphlets which can be accomplished {\em far} faster and produce results with low frequency error and high confidence (see for example \cite{rossi2017estimation,yang2018ssrw,hasan2017graphettes}).

\subsubsection{Which topological score to use?}
We believe that one of the major outstanding questions in network alignment is the design of good topological objective functions. While most measures that currently exist have been shown to correlate with interesting biological information, none have been shown to be substantially better than any other in terms of recovering relevant biology. For example, while $S^3$ is symmetric and can thus be considered a more aesthetically pleasing measure from a mathematical standpoint, it's by no means clear that it actually produces better correlations with biology than $EC$. And while graphlets have been shown to correlate with biological information \citep{GRAAL,LGRAAL,DavisPrzulj2015TopoFunction}, it is not clear that we know the best way to {\em use} them to recover the greatest amount of relevant biological information (cf. Section \ref{sec:Jurisica}, especially Table \ref{tab:Jurisica}). In general, the design of good topological objective functions is a wide-open area of research that deserves to be explored. SANA, with its speed and accuracy, is an ideal playground for exploring objective functions.

\begin{figure}[hbt]
    \centering
    \includegraphics[width=0.95\linewidth]{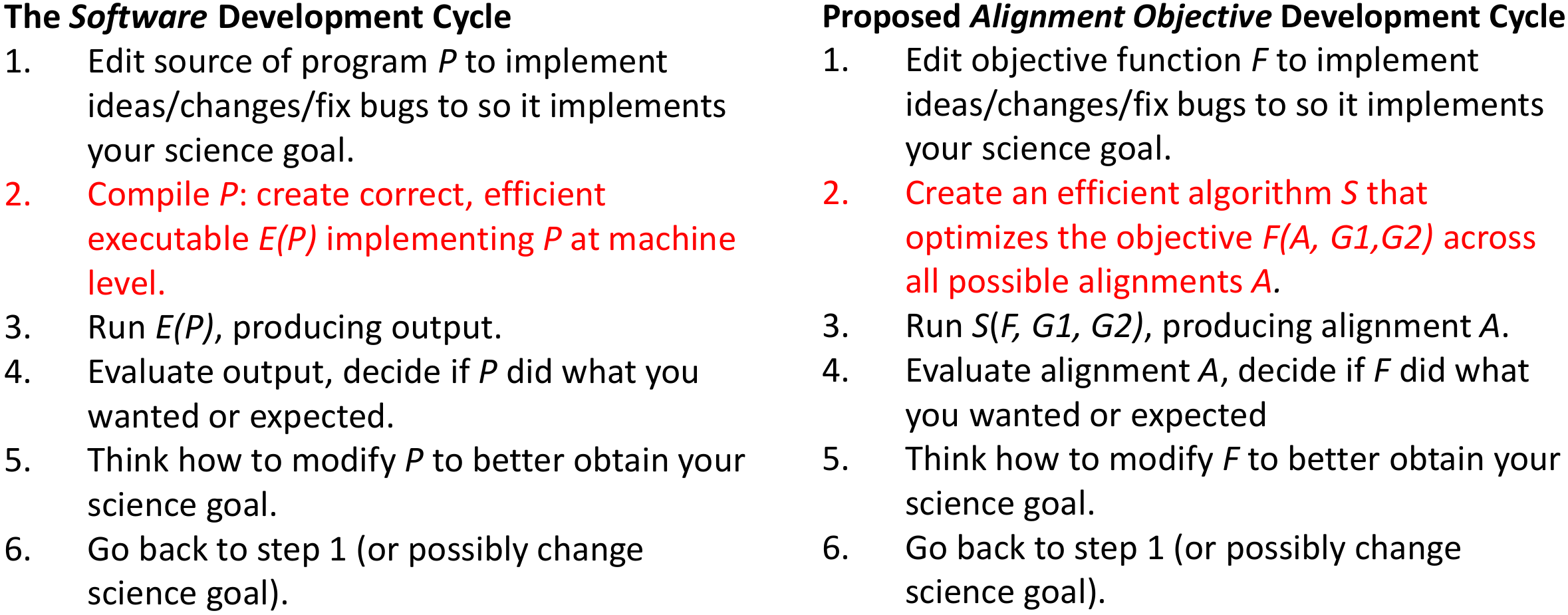}
    \caption{\small Comparison of the standard software development cycle (left), and proposed cycle for developing new objective functions for alignment (right). Red highlights the step that should be entirely automated and requiring no effort on the user's part.}
    \label{fig:DevCycle}
\end{figure}

To explain what we mean by experimenting with objective functions, consider Figure \ref{fig:DevCycle}.
There are three orthogonal components to network alignment:
{\bf (1)} a (possibly vague) scientific or informational goal $G$; {\bf (2)} an objective function $M$ created by the user that attempts to formally encode $G$; and {\bf (3)} an alignment algorithm $S$ that builds an alignment trying to optimize $M$.
In sequence alignment, the three orthogonal components are clearly delimited: the substitution/indel cost matrix encodes the goal the user wants, and tools like BLAST \citep{BLAST} quickly find (near-)optimal solutions. Practitioners can {\em use} BLAST without having to understand the details of how it works.  It is a trusted tool, like a C++ compiler is to a developer, or a linear solver to a scientist solving a linear system; practitioners iterate the familiar edit-compile-debug loop, gaining knowledge from the feedback process until they are satisfied that they have achieved their goal.
Unfortunately, this edit-compile-debug loop is virtually impossible in the network alignment arena, due to (i) the the lack of an algorithm fast enough to perform effective edit-compile-debug loops, (ii) the lack of a generally-accepted ``gold standard'' of network alignment, and (iii) the lack of a clear separation of the {\em goal}, its formalized {\em objective}, and the {\em alignment tool}. SANA fixes the first two; the third is a matter of scientific culture in the network alignment community that we hope to influence by spreading the use of SANA in conjunction with the process depicted in Figure \ref{fig:DevCycle}.

\subsubsection{Using sequence similarity as an objective---a necessary but hopefully temporary evil}
It may help here to (re-)state the obvious: the whole point of {\em network} alignment is to align networks based upon their network topology. This is a desirable goal because there is a strong belief that the topology of a network is somehow related to its function. For example, we believe that humans and chimpanzees are very close relatives, taxonomically speaking. If there is a particular protein $h_0$ in humans that performs a certain function by interacting with seven other proteins $h_1,h_2,\ldots,h_7$, then it is quite likely that there is a very similar protein $c_0$ in chimpanzees that also interacts with (close to) seven proteins $c_1,c_2,\ldots,c_7$ to perform virtually the same function. Another way of saying this is that the network topology of the protein-protein interaction networks of human and chimp are likely to be very similar in the vicinity of $h_0$ and $c_0$, respectively. As such, a natural network alignment between human and chimp should contain the ordered pairs $(h_0,c_0),(h_1,c_1),\ldots,(h_7,c_7)$. If the network of interactions around $h_0$ and $c_0$ are in fact similar, then any network alignment algorithm worth its mettle, optimizing an objective that highlights such network similarities, should include the above pairs with high likelihood.

The problem, at least in the research area of protein-protein interaciton (PPI) networks, is that the data on current PPI networks is {\em extremely} incomplete in terms of enumerating the edges in the PPI networks.  For example, as of 2018, the most complete PPI network is that of {\it S. cerevisiae}, and it may be only about 50\% complete; the human PPI network is probably less than 10\% complete \citep{vidal2016much}; other species are even far less complete.  For instance, we'd expect most mammals to have about the same number of interactions in their PPI networks, and yet the 2018 BioGRID Human network has almost 300,000 interactions, but mouse and rat have only 38,000 and 5,000 interactions listed, respectively. If Human is only 10\% complete and currenthly contains 300,000 interactions, then we may expect the complete interactome to have over 1 million interactions. By this measure, mouse and rat are at most a few percent, and well less than one percent complete, respectively. Here's the crux: if we are missing 90\% or more of the edges in most mammal PPI networks, {\em no network alignment algorithm based solely upon network topology has any hope of providing good alignments}. This is the state of affairs in PPI network alignment.

Thus, it is no surprise that virtually every network alignment algorithm currently in existence must rely on using sequence similarity information to help give network alignments that show decent functional similarity. However, {\em if network alignment is of any worth whatsoever, the use of sequence similarity should be viewed only as a {\bf temporary} crutch---a necessary evil---until such time as the interactions in PPI networks are more completely enumerated}.

On the other hand, since protein function is defined by the shape of the folded protein, and disrupting the function of a protein can be lethal, the folded structure of a protein tends to be better conserved than its sequence \citep{lesk1986response}. This in turn suggests that the network of interactions may also be better conserved than sequence. If this is the case, then network alignment may ultimately be at least as useful as sequence alignment in terms of learning about protein function. Alas, we must wait until PPI networks are far more complete than they are today to test this hypothesis.

\subsection{Search Algorithms}
Given two networks with $n_1\le n_2$ nodes, respectively, the number of possible 1-to-1 pairwise global network alignments between them is exactly \(\frac{n_2!}{(n_2-n_1)!}\). This is an {\em enormous} number; for example if the two networks each have thousands of nodes (not uncommon for protein-protein interaction networks), the the number of possible alignments can easily exceed $10^{100,000}$.  This is an enormous search space, {\em far} larger, for example, than the number of elementary particles in the known universe---which according to Wikipedia is a paltry $10^{100}$.

The task of a network alignment algorithm is to search through this enormous space of possible alignments, looking for ones that score well according to one or more of the measures described in Section \ref{sec:measures}. Since network alignment is an NP-complete problem\footnote{For those who are inclined to graph theory, the proof is trivial: finding a network alignment with an EC of exactly 1 is equivalent to solving the subgraph isomorphism problem.}, all such algorithms must use heuristics to navigate this enormous search space.
Search methods abound; several good review papers exist \citep{clark2014comparison,TijanaReview2015,milano2017extensive,guzzi2017survey}; for an extensive comparison specifically showing that SANA outperforms about a dozen of the best existing algorithms, see \cite{MamanoHayesSANA}.  SANA is virtually unique in that it was designed from the start to be able to optimize {\em any} objective function, including the objective functions introduced by other researchers; a preliminary report shows that SANA outperforms over a dozen other algorithms at optimizing their own objective functions \citep{KanneHayesArXiv}.

\subsection{Requirements of a good alignment algorithm}
We believe that, in order to be of general use, a network alignment algorithm must satisfy the following properties:

    \paragraph{Speed, if so desired.} SANA can produce better alignments in minutes that most other aligners can in hours. This is useful for many reasons: to perform test alignments; to experiment with objective functions; to perform multiple alignments of the same pair of networks in order to see which parts of the alignment, if any, come out the same each time (more on this later).
    \paragraph{High quality of results, if so desired.} SANA's primary user-tunable parameter is the amount of time the user wishes to wait. While SANA can produce better alignments in one minute on a laptop than many existing algorithms can do given hours of CPU, users can also tell SANA to spend any amount of time improving the alignment, such as 5 minutes, 3 hours, or a week. SANA generally produces better scoring alignments with longer run times, although we generally see a point of diminishing returns beyond a few hours.
    \paragraph{It should be simple to use.} By this we mean that if there are any algorithmic parameters that crucially control the quality of the result, those parameters should be tuned automatically without user input---in other words, the user should not need be an expert on the algorithm in order to understand how to use it. The primary internal parameters controlling the anneal is the temperature schedule, and by default SANA spends a minute or two automatically finding a near-optimal temperature schedule before starting the anneal. (Another algorithm called SailMCS \citep{SailMCS} also uses simulated annealing but fails to automatically determine a good temperature schedule, and so SANA produces alignments that are far superior to those of SailMCS \citep{KanneHayesArXiv}.)
    \paragraph{Providing confidence estimates on the quality of the alignment.} For example, if some set of pegs $P_1$ always end up in the same holes every time SANA is run and another set of pegs $P_2$ end up in different holes each time SANA is run, this suggests the set $P_1$ is {\em confidently} aligned, whereas we should be suspicious about the alignment of pegs in $P_2$. Few algorithms are capable of this sort of confidence testing of the alignment; SANA, on the other hand, is so fast that it is easy to look for such {\em core} alignments \citep{HGRAAL}---cf. Section \ref{sec:Jurisica}.
    \paragraph{Flexible with objective functions.} SANA has over a dozen pre-programmed objective functions that users can experiment with. In addition, users can supply SANA with externally computed similarity matrices, either node-to-node, or edge-to-edge. Finally, we have tried to make the code base of SANA clear so that anybody familiar with C++ can program new objective functions easily.
    \paragraph{Able to handle nodes that have ASCII names rather than only allowing integers as node identifiers.} To a programmer, creating a mapping between ASCII names and integers is easy. To non-programmers this is not so easy, and many aligners have the inexcusable fault of insisting that nodes are named by sequential integers. SANA does this internally but allows users to use whatever names they want to identify nodes.
    \paragraph{Available to plug in to existing popular tools such as Cytoscape.}  SANA is available in the Cytoscape App store.
    \paragraph{Able to handle multiple input graph formats.} Currently SANA only natively accepts networks in edge list format, and LEDA.gw format.  The former is a line-by-line list of edges (two nodes from the same network listed on one line), while the latter is a rather deprecated format used by an old version of LEDA \citep{LEDA}. However, we do provide a converter called {\tt createEdgeList} that outputs our edge list format given any of the following input formats: GML, XML, graphML, LEDA, CSV, and LGF.

\subsection{The value of randomness: core alignments}
SANA shares one important aspect with a few other aligners including MAGNA \citep{MAGNA,MAGNA++} and OptNetAlign \citep{OptNetAlign}: it is a randomized search algorithm. Like these other algorithms, SANA starts with a random alignment and then starts to move pegs around between holes; each time it tries to swap or move pegs around, it asks if the objective function has gotten better or not. As time progresses, the alignment gets better according to the objective function. If the objective function is an easy one to optimize, SANA will quickly find the optimal or near-optimal alignment \citep{MamanoHayesSANA,KanneHayesArXiv}; in harder cases it will simply find better-and-better solutions as it is given more time.

The fact that SANA intentionally injects randomness has some surprising positive aspects.  In particular, if there exist highly similar regions between the two networks $G_1$ and $G_2$, SANA is likely to find them and align them identically every time, despite starting with a different random alignment each time.  If there are other parts of the networks that are dissimilar and there is no obvious way to align them correctly, those regions are likely to get aligned differently each time SANA is run.  Given two regions $R_1$ in $G_1$ and $R_2$ in $G_2$, the more topologically similar $R_1$ is to $R_2$, the more likely it is that SANA will align them the same way every time it is run, independent of the randomness.  Since SANA is extremely fast, and since it has this random aspect, it is relatively painless to run SANA many times on the same pair of networks and look for pairs of nodes that are aligned together frequently. We use the term {\em core alignment} to refer to pairs of nodes that are stable across many runs of SANA; the more frequently a pair of nodes is aligned together, the more confident we are that they truly {\em belong} together according to the objective function being optimized.  So for example, if we run SANA 10 times on the same 2 networks and produce output files out0.align, out1.align, out2.align, $\ldots$, out9.align, then we can trivially measure the core frequencies on the Unix command line as follows:

\noindent{\tt \$ sort out?.align | uniq -c | sort -nr}

\noindent The first {\tt sort} puts identical lines from all 10 files side-by-side; the {\tt uniq -c} counts how many unique lines are side-by-side (thus measuring core frequency), and the final {\tt sort -nr} then sorts the aligned pairs of nodes by frequency, most frequent pairs of nodes first---that is, the most confident parts of the alignment are listed first.
Note that the output of the above command line is a list of pairs precedid by their frequency. Note in particular that, even though SANA is a 1-to-1 aligner {\em per run}, with multiple runs we can produce non-1-to-1 mappings between the two networks, along with a confidence level for each particular pair.\footnote{We are also working on functionality to produce core alignments in one run of SANA; that functionality may exist by the time this article goes to press and accessible via the command-line option ``{\tt -cores}''.}

\subsection{Limitations of SANA}
Currently, SANA aligns only two networks at a time. Each time, it produces a 1-to-1 mapping between the nodes of the smaller network to the nodes of the larger one (ie., an arrangement of pegs into holes). So technically, SANA is a global, pairwise, 1-to-1 alignment algorithm---the simplest type of global alignment algorithm. However, as we described above, SANA produces {\em good} alignments so quickly that it can be run many times on the same pair of networks in the same time it takes to run most other algorithms just once; by running SANA many times we effectively produce not only a non-1-to-1 mapping, but also a {\em confidence estimate} of each pair of nodes we output. So far as we are aware, no other algorithm produces such confidence estimates.

Furthermore, even though SANA technically aligns only 2 networks at a time, in the Appendix of this paper we describe a prototype version of multi-SANA that uses pairwise alignments to construct a multiple network alignment.

Thus, although SANA is technically only a 1-to-1 pairwise network aligner, it can effectively produce both many-to-many alignments (with confidences), and multiple alignments.

\section{Examples of usage}

\subsection{Getting started with SANA}
\begin{table*}
\begin{verbatim}
# Lines like this are comments. The Unix/Bash prompt is the dollar sign.
# First use "git" to clone the repo:
$ git clone http://github.com/waynebhayes/SANA
Cloning into 'SANA'... #git output deleted
$ cd SANA; make   # now wait a few minutes...
# Run SANA for the first time on the 2018 BioGRID networks of rat and S. pombe:
$ ./sana -fg1 networks/RNorvegicus18.el -fg2 networks/SPombe18.el
# wait while SANA computes a temperature schedule and then performs the alignment...
$ cat sana.out # look at the output file; first line is an internal
# representation of the alignment and can be ignored. 
$ head -3 sana.align # left column is a BioGRID node name from rat, right from S.pombe.
361207  2542195
316265  2541287
499382  2539901
$
\end{verbatim}
\caption{Getting started with SANA on the Unix command line. We first clone the repo from GitHub, then ``make" SANA, then run it on the two smallest BioGRID 2018 networks: {\it R. norvegicus} and {\it S. pombe}. We then look at the output file {\tt sana.out}, which contains scores and other useful information, as well as the actual alignment file {\tt sana.align}. SANA has {\it many} command-line options; type ``{\tt ../sana -h | less}'' to see a long list of them.
}
\label{tab:start}
\end{table*}
Table \ref{tab:start} contains a sequence of Unix Shell commands that will download the repo from GitHub, compile SANA, and perform your first test of SANA to ensure everything works. 

The most basic run of SANA requires the user only to specify which two networks to align; in Table \ref{tab:start} it is the 2018 BioGRID renditions of {\it Rattus norvegicus} (the common sewer rat, aka lab rat), and the single-celled yeast {\it Schizosaccharomyces pombe}. SANA defaults to using $S^3$ as the objective function, and 5 minutes as the amount of time to perform simulated annealing. Total run time is about 6--7 minutes including the initial computation of the temperature schedule, which we now describe.

Simulated annealing only works well if the temperature schedule is chosen carefully. We must start with a temperature high enough that moves are essentially random, so that even bad moves are frequently accepted (this keeps us out of local minima); and then end with a temperature low enough that only good moves are accepted (to hone in on the best local maximum once we've found its general vicinity). Empirically, we are controlling the {\it probability of accepting a bad move}, or {\tt pBad}; it must start close to 1, and end close to zero. Unfortunately there's no analytical method to compute these extremes, so the first 1-2 minutes of SANA are spent estimating the initial temperature $t_{initial}$, the final temperature $t_{final}$ that gives a pBad starting near 1 and ending near zero, along with the $t_{decay}$, the temperature decay rate that gets us from one to the other in the allotted time (5 minutes by default).

Next you will see the statement,
\noindent{\tt Start execution of SANA\_s3}
which says SANA is finally starting the anneal, optimizing {\tt s3}. After that, you'll see updates every few seconds as SANA progresses.  These updates show the update number, the elapsed time so far, the current score, some statistical theoretical values that don't concern us here, and the sampled pBad, which should start above 0.98 and end somewhere below about 1e-6.

Once SANA is finished running, there are exactly two output files (whose names can be changed with the ``-o'' option): {\tt sana.out} contains as its first (long) line an internal representation of the alignment, followed by some human-readable statistics; an example is in Table \ref{tab:sana.out}. The second file, called {\tt sana.align}, contains the actual alignment in two-column format: on each line, the left column contains a node (``peg'') from $G_1$ and the right column is the aligned node (``hole'') from $G_2$.

\begin{table*}
\caption{The {\tt sana.out} file (whose name can be changed using the {\tt -o} command-line option) contains information about the input networks (nodes, edges, connected components) and an analysis of the alignment (various measures applied to the entire alignment, and also applied to the common connected subgraphs).}
\begin{verbatim}
2018-06-15 15:21:57

G1: yeast
n    = 2390
m    = 16127
#connectedComponents = 158
Largest connectedComponents (nodes, edges) = (1994, 15819) (10, 32) (6, 11)

G2: human
n    = 9141
m    = 41456
#connectedComponents = 94
Largest connectedComponents (nodes, edges) = (8934, 41341) (5, 4) (4, 3)

Method: SANA_s3
Temperature schedule:
T_initial: 0.000316228
T_decay: 6.61993
Optimize:
weight s3: 1
Requested Execution time: 5 minutes

Actual execution time = 300.976 seconds

Random Seed: 514154230

Scores:
ec: 0.397966
mec: 1
ses: 35381
ics: 0.831563
s3: 0.368279
lccs: 0.248768
sec: 0.222913

Common subgraph:
n    = 2390
m    = 6418
#connectedComponents = 395
Largest connectedComponents (nodes, edges) = (1059, 4805) (53, 263) (48, 69)

Common connected subgraphs:
Graph  n     m      alig-edges  indu-edges  EC        ICS       S3
G1     2390  16127  6418        7718        0.397966  0.831563  0.368279
CCS_0  1059  4805   4805        5790        1.000000  0.829879  0.829879
CCS_1  53    263    263         268         1.000000  0.981343  0.981343
CCS_2  48    69     69          73          1.000000  0.945205  0.945205
CCS_3  34    68     68          70          1.000000  0.971429  0.971429
CCS_4  33    50     50          52          1.000000  0.961538  0.961538

\end{verbatim}
\label{tab:sana.out}
\end{table*}

The default objective function is $S^3$; changing the objective function is easy on the command line. For example to have SANA optimize a 50-50 combination of $EC$ and $S^3$, type

\noindent{\tt ./sana -ec 0.5 -s3 0.5 -fg1 ...}

\noindent To turn off $S^3$ entirely and perform an $EC$-only alignment, do

\noindent{\tt ./sana -s3 0 -ec 1 -fg1 ...}

\noindent To perform an alignmet that optimizes 90\% Importance as defined by HubAlign \citep{HubAlign} 5\% graphlets as used by GRAAL \citep{GRAAL}, 5\% EC, and no $S^3$, do

\noindent{\tt ./sana -s3 0 -importance 0.9 -graphlets 0.05 -ec 0.05 ...}

\noindent Note that one does not need to manually ensure that all the weights specified on the command line add to 1; if they do not, SANA will simply re-normalize them all so that they add to 1.

Similarly, the are many other objective functions defined by SANA; currently implemented ones are listed in Table \ref{tab:measures}.

\subsection{Direct comparison with other aligners}
As a part of our first publication on SANA \citep{MamanoHayesSANA}, we wanted to automate the process of directly comparing to many other existing aligners. Thus, the external source code of over a dozen existing aligners were directly incorporated into SANA so that they can be called from the SANA command line. This was done to ensure consistent calling conventions to these other aligners during our comparisons. These other methods can be called from the SANA command line using the {\tt -method} argument. In the SANA repo, these other aligners are in the directory {\tt wrappedAlgorithms}; see the online SANA documentation for more details.\footnote{If you are an author of one of these aligners and notice that SANA is not using your algorithm optimally, feel free to contact us with any corrections.} The other aligners currently incorporated into SANA are LGRAAL \citep{LGRAAL}, MAGNA++ \citep{MAGNA++}, HubAlign \citep{HubAlign}, WAVE \citep{WAVE}, NETAL \citep{NETAL}, MIGRAAL \citep{MIGRAAL}, GHOST \citep{GHOST}, PISWAP \citep{PISWAP}, OptNetAlign \citep{OptNetAlign}, SPINAL \citep{SPINAL}, GREAT \citep{GREAT}, NATALIE 2.0 \citep{NATALIE2}, GEDEVO \citep{GEDEVO}, CytoGEDEVO \citep{CytoGEDEVO}, BEAMS \citep{BEAMS}, HGRAAL \citep{HGRAAL}, PINALOG \citep{phan2012pinalog}.

\begin{table*}
    \caption{Measures accepted by SANA on the command line. Note that ``Name'' means ``command-line option'', so for example to give {\tt ec} a weight of 0.5, use ``{\tt -ec 0.5}'' on the SANA command line.}
    \centering
    \begin{tabular}{|ll|}
    \hline
         Name & Description \\
    \hline
         {\tt s3} & Symmetric Substructure Score \citep{MAGNA} \\
         {\tt ec} & Edge Coverage/Correspondence/Correctness \citep{GRAAL} \\
         {\tt ics} & Induced Conserved Structure \citep{GHOST} \\
         {\tt graphlet} & Orbit Degree Vector (ODV) Similarity \citep{Tijana2008,GRAAL} \\
         {\tt graphletlgraal} & LGRAAL-normalization of ODV sim \citep{LGRAAL}\\
         {\tt go} & Mean ResnikMax GO similarity \citep{resnik1995using,GOterms}\\
         {\tt NetGO} & Network-alignment-based GO similarity \citep{hayes2017sana}\\
         {\tt wec} & Weighted EC \citep{WAVE} \\
         {\tt esim} & External file defining node-pair similarities \\
         {\tt sequence} & BLANT-based sequence similarities \citep{BLAST} \\
         {\tt lccs} & Largest Common Connected Subgraph \citep{GRAAL}\\
         {\tt nc} & Node Correctness (if known, defines the exact alignment) \\
         {\tt spc} & Shortest Path Conservation \citep{MamanoHayesSANA} \\
         {\tt edgeCount} & degree difference \\
         {\tt edgeDensity} & relative degree difference \\
         {\tt importance} & HubAlign's Importance \citep{HubAlign} \\
         {\tt nodeDensity} & local node density \\
         {\tt ewec} & External edge-based similarity matrix, eg., edge-graphlet similarity\citep{GREAT} \\
         {\tt sequence} & BLAST bit scores based on protein sequence similarity \citep{BLAST}\\
    \hline
    \end{tabular}
    \label{tab:measures}
\end{table*}

\section{An example of objective function experimentation}

As shown in Figure \ref{fig:DevCycle}, SANA can be used to experiment with objective functions; we believe that such experimentation is one of the most important but apparently under-appreciated aspects of the science of network alignment.  Here we describe one such experiment with a very well-defined scientific goal.

\subsection{Gene--microRNA networks}\label{sec:Jurisica}
\begin{table*}
    \caption{Table of results when testing various objective functions (leftmost column) for their ability to correctly align genes-to-genes, and mRNAs-to-mRNAs, when aligning a pair of gene-mNRA networks
    \citep{tokar2017mirdip}. Objectives tested were $EC$ \citep{GRAAL}, $S^3$ \citep{MAGNA}, Importance \citep{HubAlign}, graphlet \citep{Tijana2008,GRAAL}, and graphlet-LGRAAL \citep{LGRAAL}. The columns are as follows. {\bf pairs}: total number of pairs of nodes aligned in all 1,000 network pairs that were run 5 times each. {\bf 2*Gene}: number of pairs in which a gene was correctly aligned to another gene. {\bf mix}: number of pairs in which a gene in one network was aligned to an mRNA in the other. {\bf 2*RNA}: number of pairs in which an mRNA was aligned to another mRNA. {\bf coreFreq(XY)$>1$}: the number of aligned pairs that had a core frequency greater than 1 (indicating the objective function strongly prefers to align this pair of nodes together) for type-pairs GG, MG, and MM.
    }
    \centering
    \begin{tabular}{|l||rrrr|rrr|}
\hline
{\bf 1 minute runs}  &&&&&&&\\
\hline
objective & pairs  &  2*Gene  &   mix     &  2*RNA & coreFreq(GG)$>1$ &  coreFreq(MG)$>1$ & coreFreq(MM)$>1$ \\
\hline
ec  &    30424880 & 29953074&   198792  &273014 &    1268806  &   570 & 3169 \\
s3   &   30424880 & 29986047 &  284470  &154363 & 1093307 & 2947 & 688 \\
importance &  30241594 & 25434876  & 4658345 &148373 & 651969 & 114137 & 1386 \\
graphlet-GRAAL&  30424880 & 24109670  & 6176510 &138700 & 1902554 & 449738 & 17331 \\
graphlet-LGRAAL &  30424880 & 23056815  & 7305611   &62454     & 1718519 & 584735 & 7086 \\
\hline\hline
{\bf 4 minute runs} &&&&&&&\\
\hline
objective & pairs  &  2*Gene  &   mix     &  2*RNA & coreFreq(GG)$>1$ &  coreFreq(MG)$>1$  & coreFreq(MM)$>1$ \\
\hline
ec   &   30424880 & 30055465  & 97811   &271604  &     1245103      &   208             & 5908 \\
s3   &   30424880 & 29953309  & 283313  &188258  & 1092319 & 3779 & 1508 \\
importance &  30292547 & 25473995  & 4669942 &148610  & 652830 & 114815 & 1621 \\
graphlet-GRAAL&  30424880 & 24104880 &  6180836 &139164 & 2208583 & 502806 & 25308 \\  
graphlet-LGRAAL &  30424880 & 23051615  & 7310109 &63156  & 2090416 & 692752 & 10504 \\
\hline
    \end{tabular}
    \label{tab:Jurisica}
\end{table*}

Consider a set of gene-microRNA (mRNA) networks \citep{tokar2017mirdip}, one network for each species. These networks are bipartite, meaning that genes interact with microRNAs, but neither genes nor microRNAs interact with their own type. Thus, when aligning two gene-mRNA networks, we wish to align genes from one network to genes in the other, and mRNAs in one to mRNAs in the other, but we should never align a gene to an mRNA or vice-versa. In essence, the nodes have two {\it types}, and we must provide a type-specific network alignment.

At first, SANA did not have the functionality to provide a typed-node alignment.\footnote{It does now, using the {\tt -nodes-have-types} argument, in which case we assume that the first column in the edge list is one type, and the second column is the other type. Only two types are supported at the moment.} The question was: how do the various topological objective functions compare in their ability to {\em automatically} align types correctly, given that typing is not {\em enforced} by the alignment algorithm?

Referring to Figure \ref{fig:DevCycle}, the scientific goal is clear: {\bf maximize the fraction of nodes that are aligned to like-type nodes in the other network}. The question is now, {\it which topological objective function best achieves this scientific goal?}

We received 535 networks directly from one of the authors of \cite{tokar2017mirdip}. We chose 1,000 pairs of networks at random out of the ${535 \choose 2}=142,845$ possible pairs of networks. For each pair of networks, we tested the following objective functions for their ability to correctly align nodes of like type to each other when this was not enforced: $EC$, $S^3$, Importance \citep{HubAlign}, GRAAL-type graphlet orbit signatures \citep{Tijana2008,GRAAL}, and LGRAAL-type graphlet orbit signatures \citep{LGRAAL}. To further test the dependence on runtime, we ran SANA on all the above objectives for all 1,000 networks for runtimes of 1 and 4 minutes. Finally, to look at the frequency of core alignments, we performed each of the above pairs 5 times each.
The results are in Table \ref{tab:Jurisica}.

One column of great interest is the ``mix'' column, which counts the number of times, out of the approximately 30 million pairs of aligned nodes,  in which a gene from one network was aligned to an mRNA in the other network---which is the kind of mis-typed node alignment we are trying to avoid. The rows are sorted best-to-worst by this measure, in each of the 1-minute and 4-minute sub-tables. As we can see, the $EC$ objective scores best at avoiding this kind of mis-typed alignment. In the 1-minute runs, $EC$ aligns unlike typed node-pairs in only 0.65\% of cases; $S^3$ is a close second, mis-typing just under 1\% of the aligned pairs of nodes.  In contrast, HubAlign's Importance measure \citep{HubAlign} is almost 20 times worse in terms of incorrectly aligning nodes of different types, doing so in about 15\% of aligned pairs of nodes, while both graphlet measures fare the worst, aligning unlike-type nodes in over 20\% of cases.

Even more interesting is the 4 minute runs, in which $EC$ cuts its mis-typed node alignment in half, down to about 0.3\% of aligned pairs, while all other measures fail to improve their ``mix'' column with the longer runtime. 

Recall that if SANA aligns the same pair of nodes together in more than one run, we say that pair is in the {\em core} alignment, because the objective function is unlikely to align two nodes together more than once by chance. Another column of great interest is thus the {\bf coreFreq(MG)$>1$} column, which tells us how frequently the objective function seems to {\em strongly} prefer mis-aligning a pair of nodes of different types. Again we see that the $EC$ measure is by far the best measure by this criterion: in the 1 minute runs, only 570 mistyped pairs appear out of 30 million (about 2 per 100,000 pairs), while the 4 minute runs cut that ``error rate'' in half, suggesting that longer runs will do a better job of correctly aligning types. Meanwhile, $S^3$ does 10x worse at 1 minute and gets {\em more} bad in the 4 minute runs, while importance and both graphlet measures misalign orders of magnitude more typed pairs, presenting a strong preference for misaligning nodes in about 1--2\% of pairs.

{\bf We conclude that the $EC$ measure is, {\em by far}, the best available objective function for this particular purpose among those we tested.}
For the moment we do not hypothesize why this is the case, but empirically the result seems iron-clad.
While we agree that the $S^3$ measure is mathematically more aesthetically pleasing and would seem to be a better measure intuitively, for this particular purpose $EC$ seems to work better.
The author finds the poor performance of graphlet-based measures particularly surprising, since the author is a strong believer that graphlets are a useful tool for network analysis (see for example \cite{hasan2017graphettes})---and graphlets have certainly demonstrated their value in other contexts \citep{DavisPrzulj2015TopoFunction,Przulj2014HiddenLanguage}.  However, these results suggest that perhaps orbit degree signatures as they are currently defined \citep{Tijana2008,GRAAL,LGRAAL} may not be the best way to leverage graphlet-based information in the context of global pairwise network alignment.

\section{Conclusion}
We have described the use of SANA \citep{MamanoHayesSANA}, the {\it Simulated Annealing Network Aligner}, in the context of the pairwise 1-to-1 global alignment of biological networks.  SANA provides many advantages over the many other aligners currently available: as a search algorithm, it is lightning fast, producing well-scoring alignments in minutes rather than hours; it provides a large array of objective functions users may wish to experiment with, as well as the facility to add more objectives in the future; it does not require the user to know much about the internal workings of the aligner in order to use it; and it is well on the way towards being fully integrated into popular network analysis tools such as Cytoscape.

We have introduced the concept of {\it objective function experimentation} (cf. Figure \ref{fig:DevCycle} and Section \ref{sec:Jurisica}), which we believe is at the core of future developments in network alignment. SANA's speed and effectiveness makes it the ideal aligner to implement the process depicted in Figure \ref{fig:DevCycle}.

\section*{Appendix}
A prototype of a multiple-network-alignment version of SANA is available in the SANA GitHub repo. Simply re-compile SANA with the {\tt -DWEIGHTED} option on the command line (see the {\tt Makefile}), and the consult the Bourne shell script {\tt multi-pairwise.sh}; running it without any arguments provides a short help message.

Questions about SANA, comments, or feature requests should be directed to the author at {\tt whayes@uci.edu}.

\bibliographystyle{natbib}
\bibliography{wayne-all}{}

\begin{thebibliography}{}

\bibitem[Alada\u{g} and Erten(2013)Alada\u{g} and Erten]{SPINAL}
Alada\u{g}, A.~E. and Erten, C. (2013).
\newblock Spinal: scalable protein interaction network alignment.
\newblock {\em Bioinformatics\/}, {\bf 29}(7), 917--924.

\bibitem[Alkan and Erten(2014)Alkan and Erten]{BEAMS}
Alkan, F. and Erten, C. (2014).
\newblock Beams: backbone extraction and merge strategy for the global
  many-to-many alignment of multiple ppi networks.
\newblock {\em Bioinformatics\/}, {\bf 30}(4), 531--539.

\bibitem[Ashburner {\em et~al.}(2000)Ashburner, Ball, Blake, Botstein, Butler,
  Cherry, Davis, Dolinski, Dwight, Eppig, Harris, Hill, Issel-Tarver,
  Kasarskis, Lewis, Matese, Richardson, Ringwald, Rubin, and Sherlock]{GOterms}
Ashburner, M., Ball, C.~A., Blake, J.~A., Botstein, D., Butler, H., Cherry,
  J.~M., Davis, A.~P., Dolinski, K., Dwight, S.~S., Eppig, J.~T., Harris,
  M.~A., Hill, D.~P., Issel-Tarver, L., Kasarskis, A., Lewis, S., Matese,
  J.~C., Richardson, J.~E., Ringwald, M., Rubin, G.~M., and Sherlock, G.
  (2000).
\newblock {Gene Ontology: tool for the unification of biology}.
\newblock {\em Nature Genetics\/}, {\bf 25}(1), 25--29.

\bibitem[Camacho {\em et~al.}(2009)Camacho, Coulouris, Avagyan, Ma,
  Papadopoulos, Bealer, and Madden]{BLAST}
Camacho, C., Coulouris, G., Avagyan, V., Ma, N., Papadopoulos, J.~S., Bealer,
  K., and Madden, T.~L. (2009).
\newblock Blast+: architecture and applications.
\newblock {\em BMC Bioinformatics\/}, {\bf 10}, 421.

\bibitem[Chatr-Aryamontri {\em et~al.}(2017)Chatr-Aryamontri, Oughtred,
  Boucher, Rust, Chang, Kolas, O'Donnell, Oster, Theesfeld, Sellam, {\em
  et~al.}]{chatr2017biogrid}
Chatr-Aryamontri, A., Oughtred, R., Boucher, L., Rust, J., Chang, C., Kolas,
  N.~K., O'Donnell, L., Oster, S., Theesfeld, C., Sellam, A., {\em et~al.}
  (2017).
\newblock The biogrid interaction database: 2017 update.
\newblock {\em Nucleic acids research\/}, {\bf 45}(D1), D369--D379.

\bibitem[Chen and Rajewsky(2007)Chen and Rajewsky]{chen2007evolution}
Chen, K. and Rajewsky, N. (2007).
\newblock The evolution of gene regulation by transcription factors and
  micrornas.
\newblock {\em Nature reviews. Genetics\/}, {\bf 8}(2), 93.

\bibitem[Chindelevitch {\em et~al.}(2013)Chindelevitch, Ma, Liao, and
  Berger]{PISWAP}
Chindelevitch, L., Ma, C.-Y., Liao, C.-S., and Berger, B. (2013).
\newblock Optimizing a global alignment of protein interaction networks.
\newblock {\em Bioinformatics\/}, {\bf 29}(21), 2765--2773.

\bibitem[Clark and Kalita(2014)Clark and Kalita]{clark2014comparison}
Clark, C. and Kalita, J. (2014).
\newblock A comparison of algorithms for the pairwise alignment of biological
  networks.
\newblock {\em Bioinformatics\/}, {\bf 30}(16), 2351--2359.

\bibitem[Clark and Kalita(2015)Clark and Kalita]{OptNetAlign}
Clark, C. and Kalita, J. (2015).
\newblock A multiobjective memetic algorithm for ppi network alignment.
\newblock {\em Bioinformatics\/}, {\bf 31}(12), 1988--1998.

\bibitem[Consortium(2008)Consortium]{GO}
Consortium, T. G.~O. (2008).
\newblock The gene ontology project in 2008.
\newblock {\em Nucleic Acids Research\/}, {\bf 36}(suppl 1), D440--D444.

\bibitem[Cook(1971)Cook]{cook1971complexity}
Cook, S.~A. (1971).
\newblock The complexity of theorem-proving procedures.
\newblock In {\em Proceedings of the third annual ACM symposium on Theory of
  computing\/}, pages 151--158. ACM.

\bibitem[Crawford and Milenkovi{\'c}(2015)Crawford and Milenkovi{\'c}]{GREAT}
Crawford, J. and Milenkovi{\'c}, T. (2015).
\newblock Great: graphlet edge-based network alignment.
\newblock In {\em Bioinformatics and Biomedicine (BIBM), 2015 IEEE
  International Conference on\/}, pages 220--227. IEEE.

\bibitem[Davidson(2010)Davidson]{davidson2010regulatory}
Davidson, E.~H. (2010).
\newblock {\em The regulatory genome: gene regulatory networks in development
  and evolution\/}.
\newblock Academic press, USA.

\bibitem[Davis {\em et~al.}(2015)Davis, Yavero{\u{g}}lu, Malod-Dognin,
  Stojmirovic, and Pr{\v{z}}ulj]{DavisPrzulj2015TopoFunction}
Davis, D., Yavero{\u{g}}lu, {\"O}.~N., Malod-Dognin, N., Stojmirovic, A., and
  Pr{\v{z}}ulj, N. (2015).
\newblock Topology-function conservation in protein--protein interaction
  networks.
\newblock {\em Bioinformatics\/}, {\bf 31}(10), 1632--1639.

\bibitem[El-Kebir {\em et~al.}(2011)El-Kebir, Heringa, and Klau]{NATALIE2}
El-Kebir, M., Heringa, J., and Klau, G.~W. (2011).
\newblock Lagrangian relaxation applied to sparse global network alignment.
\newblock In {\em IAPR International Conference on Pattern Recognition in
  Bioinformatics\/}, pages 225--236. Springer.

\bibitem[Faisal {\em et~al.}(2015a)Faisal, Meng, Crawford, and
  Milenkovi{\'c}]{faisal2015post}
Faisal, F.~E., Meng, L., Crawford, J., and Milenkovi{\'c}, T. (2015a).
\newblock The post-genomic era of biological network alignment.
\newblock {\em EURASIP Journal on Bioinformatics and Systems Biology\/}, {\bf
  2015}(1), 3.

\bibitem[Faisal {\em et~al.}(2015b)Faisal, Meng, Crawford, and
  Milenkovi{\'c}]{TijanaReview2015}
Faisal, F.~E., Meng, L., Crawford, J., and Milenkovi{\'c}, T. (2015b).
\newblock The post-genomic era of biological network alignment.
\newblock {\em EURASIP Journal on Bioinformatics and Systems Biology\/}, {\bf
  2015}(1), 1.

\bibitem[Farazi {\em et~al.}(2013)Farazi, Hoell, Morozov, and
  Tuschl]{farazi2013micrornas}
Farazi, T.~A., Hoell, J.~I., Morozov, P., and Tuschl, T. (2013).
\newblock Micrornas in human cancer.
\newblock In {\em MicroRNA Cancer Regulation\/}, pages 1--20. Springer,
  Germany.

\bibitem[Fiehn(2002)Fiehn]{fiehn2002metabolomics}
Fiehn, O. (2002).
\newblock Metabolomics-the link between genotypes and phenotypes.
\newblock In {\em Functional Genomics\/}, pages 155--171. Springer, Germany.

\bibitem[Garey and Johnson(1979)Garey and Johnson]{GareyJohnson}
Garey, M. and Johnson, D. (1979).
\newblock {\em Computers and Intractability: A Guide to the Theory of
  NP-Completeness\/}.
\newblock New York: W.H. Freeman, New York.

\bibitem[Guzzi and Milenkovi{\'c}(2017)Guzzi and
  Milenkovi{\'c}]{guzzi2017survey}
Guzzi, P.~H. and Milenkovi{\'c}, T. (2017).
\newblock Survey of local and global biological network alignment: the need to
  reconcile the two sides of the same coin.
\newblock {\em Briefings in bioinformatics\/}, page bbw132.

\bibitem[Hasan {\em et~al.}(2017)Hasan, Chung, and Hayes]{hasan2017graphettes}
Hasan, A., Chung, P.-C., and Hayes, W. (2017).
\newblock Graphettes: Constant-time determination of graphlet and orbit
  identity including (possibly disconnected) graphlets up to size 8.
\newblock {\em PloS one\/}, {\bf 12}(8), e0181570.

\bibitem[Hashemifar and Xu(2014)Hashemifar and Xu]{HubAlign}
Hashemifar, S. and Xu, J. (2014).
\newblock {HubAlign: an accurate and efficient method for global alignment of
  protein–protein interaction networks}.
\newblock {\em Bioinformatics\/}, {\bf 30}(17), i438--i444.

\bibitem[Hayes and Mamano(2017)Hayes and Mamano]{hayes2017sana}
Hayes, W.~B. and Mamano, N. (2017).
\newblock Sana netgo: a combinatorial approach to using gene ontology (go)
  terms to score network alignments.
\newblock {\em Bioinformatics\/}, {\bf 34}(8), 1345--1352.

\bibitem[Ho\v{c}evar and Dem\v{s}ar(2014)Ho\v{c}evar and Dem\v{s}ar]{ORCA}
Ho\v{c}evar, T. and Dem\v{s}ar, J. (2014).
\newblock {A combinatorial approach to graphlet counting}.
\newblock {\em Bioinformatics\/}, {\bf 30}(4), 559--565.

\bibitem[Ibragimov {\em et~al.}(2013)Ibragimov, Malek, Guo, and
  Baumbach]{GEDEVO}
Ibragimov, R., Malek, M., Guo, J., and Baumbach, J. (2013).
\newblock Gedevo: an evolutionary graph edit distance algorithm for biological
  network alignment.
\newblock In {\em OASIcs-OpenAccess Series in Informatics\/}, volume~34.
  Schloss Dagstuhl-Leibniz-Zentrum fuer Informatik.

\bibitem[Jaenicke and Helmreich(2012)Jaenicke and
  Helmreich]{jaenicke2012protein}
Jaenicke, R. and Helmreich, E. (2012).
\newblock {\em Protein-protein interactions\/}, volume~23.
\newblock Springer Science \& Business Media, Germany.

\bibitem[Junker and Schreiber(2011)Junker and Schreiber]{junker2011analysis}
Junker, B.~H. and Schreiber, F. (2011).
\newblock {\em Analysis of biological networks\/}, volume~2.
\newblock John Wiley \& Sons, USA.

\bibitem[Kanne and Hayes(2017)Kanne and Hayes]{KanneHayesArXiv}
Kanne, D.~P. and Hayes, W.~B. (2017).
\newblock Sana: separating the search algorithm from the objective function in
  biological network alignment, part 1: Search.

\bibitem[Karlebach and Shamir(2008)Karlebach and
  Shamir]{karlebach2008modelling}
Karlebach, G. and Shamir, R. (2008).
\newblock Modelling and analysis of gene regulatory networks.
\newblock {\em Nature reviews. Molecular cell biology\/}, {\bf 9}(10), 770.

\bibitem[Kotlyar {\em et~al.}(2015)Kotlyar, Pastrello, Sheahan, and
  Jurisica]{kotlyar2015integrated}
Kotlyar, M., Pastrello, C., Sheahan, N., and Jurisica, I. (2015).
\newblock Integrated interactions database: tissue-specific view of the human
  and model organism interactomes.
\newblock {\em Nucleic acids research\/}, {\bf 44}(D1), D536--D541.

\bibitem[Kuchaiev and Pr\v{z}ulj(2011)Kuchaiev and Pr\v{z}ulj]{MIGRAAL}
Kuchaiev, O. and Pr\v{z}ulj, N. (2011).
\newblock Integrative network alignment reveals large regions of global network
  similarity in yeast and human.
\newblock {\em BIOINFORMATICS\/}, {\bf 27}, 1390--1396.

\bibitem[Kuchaiev {\em et~al.}(2010)Kuchaiev, Milenkovi{\'c}, Memi{\v
  s}evi{\'c}, Hayes, and Pr\v{z}ulj]{GRAAL}
Kuchaiev, O., Milenkovi{\'c}, T., Memi{\v s}evi{\'c}, V., Hayes, W., and
  Pr\v{z}ulj, N. (2010).
\newblock Topological network alignment uncovers biological function and
  phylogeny.
\newblock {\em Journal of The Royal Society Interface\/}, {\bf 7}(50),
  1341--1354.

\bibitem[Larsen {\em et~al.}(2016)Larsen, Alk{\ae}rsig, Ditzel, Jurisica,
  Alcaraz, and Baumbach]{SailMCS}
Larsen, S.~J., Alk{\ae}rsig, F.~G., Ditzel, H.~J., Jurisica, I., Alcaraz, N.,
  and Baumbach, J. (2016).
\newblock A simulated annealing algorithm for maximum common edge subgraph
  detection in biological networks.
\newblock In {\em Proceedings of the 2016 on Genetic and Evolutionary
  Computation Conference\/}, pages 341--348. ACM.

\bibitem[Lesk and Chothia(1986)Lesk and Chothia]{lesk1986response}
Lesk, A. and Chothia, C. (1986).
\newblock The response of protein structures to amino-acid sequence changes.
\newblock {\em Phil. Trans. R. Soc. Lond. A\/}, {\bf 317}(1540), 345--356.

\bibitem[Malek {\em et~al.}(2016)Malek, Ibragimov, Albrecht, and
  Baumbach]{CytoGEDEVO}
Malek, M., Ibragimov, R., Albrecht, M., and Baumbach, J. (2016).
\newblock Cytogedevo—global alignment of biological networks with cytoscape.
\newblock {\em Bioinformatics\/}, {\bf 32}(8), 1259--1261.

\bibitem[Malod-Dognin and Pr\v{z}ulj(2015)Malod-Dognin and Pr\v{z}ulj]{LGRAAL}
Malod-Dognin, N. and Pr\v{z}ulj, N. (2015).
\newblock L-graal: Lagrangian graphlet-based network aligner.
\newblock {\em Bioinformatics\/}.

\bibitem[Mamano and Hayes(2017)Mamano and Hayes]{MamanoHayesSANA}
Mamano, N. and Hayes, W.~B. (2017).
\newblock Sana: Simulated annealing far outperforms many other search
  algorithms for biological network alignment.
\newblock {\em Bioinformatics (Oxford, England)\/}, {\bf 33}, 2156–2164.

\bibitem[Mehlhorn and Naher(1999)Mehlhorn and Naher]{LEDA}
Mehlhorn, K. and Naher, S. (1999).
\newblock {\em Leda: A platform for combinatorial and geometric computing.}
\newblock Cambridge University Press, United Kingdom.

\bibitem[Milano {\em et~al.}(2017)Milano, Guzzi, Tymofieva, Xu, Hess, Veltri,
  and Cannataro]{milano2017extensive}
Milano, M., Guzzi, P.~H., Tymofieva, O., Xu, D., Hess, C., Veltri, P., and
  Cannataro, M. (2017).
\newblock An extensive assessment of network alignment algorithms for
  comparison of brain connectomes.
\newblock {\em BMC bioinformatics\/}, {\bf 18}(6), 235.

\bibitem[Milenkovi\'{c} and Pr\v{z}ulj(2008)Milenkovi\'{c} and
  Pr\v{z}ulj]{Tijana2008}
Milenkovi\'{c}, T. and Pr\v{z}ulj, N. (2008).
\newblock Uncovering biological network function via graphlet degree
  signatures.
\newblock {\em Cancer Inform.}, {\bf 6}(Epub 2008 Apr 14), 257--273.

\bibitem[Milenkovi\'{c} {\em et~al.}(2010)Milenkovi\'{c}, Ng, Hayes, and
  Pr\v{z}ulj]{HGRAAL}
Milenkovi\'{c}, T., Ng, W.~L., Hayes, W., and Pr\v{z}ulj, N. (2010).
\newblock Optimal network alignment with graphlet degree vectors.
\newblock {\em Cancer Informatics\/}, {\bf 9}, 121--137.

\bibitem[Neyshabur {\em et~al.}(2013)Neyshabur, Khadem, Hashemifar, and
  Arab]{NETAL}
Neyshabur, B., Khadem, A., Hashemifar, S., and Arab, S.~S. (2013).
\newblock Netal: a new graph-based method for global alignment of
  protein–protein interaction networks.
\newblock {\em Bioinformatics\/}, {\bf 29}(13), 1654--1662.

\bibitem[Patro and Kingsford(2012)Patro and Kingsford]{GHOST}
Patro, R. and Kingsford, C. (2012).
\newblock Global network alignment using multiscale spectral signatures.
\newblock {\em Bioinformatics\/}, {\bf 28}(23), 3105--3114.

\bibitem[Phan and Sternberg(2012)Phan and Sternberg]{phan2012pinalog}
Phan, H.~T. and Sternberg, M.~J. (2012).
\newblock Pinalog: a novel approach to align protein interaction
  networks---implications for complex detection and function prediction.
\newblock {\em Bioinformatics\/}, {\bf 28}(9), 1239--1245.

\bibitem[Prescott(2012)Prescott]{prescott2012cell}
Prescott, D.~M. (2012).
\newblock {\em Cell Biology A Comprehensive Treatise V3: Gene Expression: The
  Production of RNA's\/}, volume~3.
\newblock Elsevier, Amsterdam-London-New York-Oxford-Paris-Shannon-Tokyo.

\bibitem[Pr\v{z}ulj {\em et~al.}(2004a)Pr\v{z}ulj, Wigle, and
  Jurisica]{Przulj03}
Pr\v{z}ulj, N., Wigle, D., and Jurisica, I. (2004a).
\newblock Functional topology in a network of protein interactions.
\newblock {\em Bioinformatics\/}, {\bf 20}(3), 340--348.

\bibitem[Pr\v{z}ulj {\em et~al.}(2004b)Pr\v{z}ulj, Corneil, and
  Jurisica]{Przulj2004}
Pr\v{z}ulj, N., Corneil, D.~G., and Jurisica, I. (2004b).
\newblock Modeling interactome: scale-free or geometric?
\newblock {\em Bioinformatics\/}, {\bf 20}(18), 3508--3515.

\bibitem[Resnik(1995)Resnik]{resnik1995using}
Resnik, P. (1995).
\newblock Using information content to evaluate semantic similarity in a
  taxonomy.
\newblock {\em arXiv preprint cmp-lg/9511007\/}.

\bibitem[Rossi {\em et~al.}(2017)Rossi, Zhou, and Ahmed]{rossi2017estimation}
Rossi, R.~A., Zhou, R., and Ahmed, N.~K. (2017).
\newblock Estimation of graphlet statistics.
\newblock {\em arXiv preprint arXiv:1701.01772\/}.

\bibitem[Saraph and Milenkovi{\'c}(2014)Saraph and Milenkovi{\'c}]{MAGNA}
Saraph, V. and Milenkovi{\'c}, T. (2014).
\newblock {MAGNA}: maximizing accuracy in global network alignment.
\newblock {\em Bioinformatics\/}, {\bf 30}(20), 2931--2940.

\bibitem[Sporns(2010)Sporns]{sporns2010networks}
Sporns, O. (2010).
\newblock {\em Networks of the Brain\/}.
\newblock MIT press, USA.

\bibitem[Sun {\em et~al.}(2015)Sun, Crawford, Tang, and Milenkovi\`{c}]{WAVE}
Sun, Y., Crawford, J., Tang, J., and Milenkovi\`{c}, T. (2015).
\newblock Simultaneous optimization of both node and edge conservation in
  network alignment via {WAVE}.
\newblock In M.~Pop and H.~Touzet, editors, {\em Algorithms in
  Bioinformatics\/}, volume 9289 of {\em Lecture Notes in Computer Science\/},
  pages 16--39. Springer Berlin Heidelberg, Germany.

\bibitem[Tokar {\em et~al.}(2017)Tokar, Pastrello, Rossos, Abovsky, Hauschild,
  Tsay, Lu, and Jurisica]{tokar2017mirdip}
Tokar, T., Pastrello, C., Rossos, A.~E., Abovsky, M., Hauschild, A.-C., Tsay,
  M., Lu, R., and Jurisica, I. (2017).
\newblock mirdip 4.1—integrative database of human microrna target
  predictions.
\newblock {\em Nucleic acids research\/}, {\bf 46}(D1), D360--D370.

\bibitem[Van~El {\em et~al.}(2013)Van~El, Cornel, Borry, Hastings, Fellmann,
  Hodgson, Howard, Cambon-Thomsen, Knoppers, Meijers-Heijboer, {\em
  et~al.}]{van2013whole}
Van~El, C.~G., Cornel, M.~C., Borry, P., Hastings, R.~J., Fellmann, F.,
  Hodgson, S.~V., Howard, H.~C., Cambon-Thomsen, A., Knoppers, B.~M.,
  Meijers-Heijboer, H., {\em et~al.} (2013).
\newblock Whole-genome sequencing in health care: recommendations of the
  european society of human genetics.
\newblock {\em European Journal of Human Genetics\/}, {\bf 21}(6), 580.

\bibitem[Vidal(2016)Vidal]{vidal2016much}
Vidal, M. (2016).
\newblock How much of the human protein interactome remains to be mapped?

\bibitem[Vijayan and Milenkovi{\'c}(2017)Vijayan and
  Milenkovi{\'c}]{vijayan2017aligning}
Vijayan, V. and Milenkovi{\'c}, T. (2017).
\newblock Aligning dynamic networks with dynawave.
\newblock {\em Bioinformatics\/}, {\bf 34}(10), 1795--1798.

\bibitem[Vijayan {\em et~al.}(2015)Vijayan, Saraph, and
  Milenkovi{\'c}]{MAGNA++}
Vijayan, V., Saraph, V., and Milenkovi{\'c}, T. (2015).
\newblock Magna++: Maximizing accuracy in global network alignment via both
  node and edge conservation.
\newblock {\em Bioinformatics\/}, {\bf 31}(14), 2409--2411.

\bibitem[Von~Mering {\em et~al.}(2002)Von~Mering, Krause, Snel, Cornell, {\em
  et~al.}]{falsepositives}
Von~Mering, C., Krause, R., Snel, B., Cornell, M., {\em et~al.} (2002).
\newblock Comparative assessment of large-scale data sets of protein-protein
  interactions.
\newblock {\em Nature\/}, {\bf 417}(6887), 399.

\bibitem[Williamson and Sutcliffe(2010)Williamson and
  Sutcliffe]{williamson2010protein}
Williamson, M.~P. and Sutcliffe, M.~J. (2010).
\newblock Protein--protein interactions.

\bibitem[Yang {\em et~al.}(2018)Yang, Lyu, Li, Zhao, and Xu]{yang2018ssrw}
Yang, C., Lyu, M., Li, Y., Zhao, Q., and Xu, Y. (2018).
\newblock Ssrw: A scalable algorithm for estimating graphlet statistics based
  on random walk.
\newblock In {\em International Conference on Database Systems for Advanced
  Applications\/}, pages 272--288. Springer.

\bibitem[Yavero{\u{g}}lu {\em et~al.}(2014)Yavero{\u{g}}lu, Malod-Dognin,
  Davis, Levnajic, Janjic, Karapandza, Stojmirovic, and
  Pr{\v{z}}ulj]{Przulj2014HiddenLanguage}
Yavero{\u{g}}lu, {\"O}.~N., Malod-Dognin, N., Davis, D., Levnajic, Z., Janjic,
  V., Karapandza, R., Stojmirovic, A., and Pr{\v{z}}ulj, N. (2014).
\newblock Revealing the hidden language of complex networks.
\newblock {\em Scientific reports\/}, {\bf 4}, 4547.

\end{thebibliography}

\end{document}

\firstpage{1}

\title[SANA: Separating search from objective, Part 1 Supplementary Material]{SANA: separating the search algorithm from the objective function in biological network alignment, Part 1: Search--Supplementary Material}
\author[Kanne, Hayes]{Dillon Kanne, Wayne B. Hayes\footnote{to whom correspondence should be addressed ({\tt whayes@uci.edu})}\;}
\address{Department of Computer Science, University of California, Irvine CA 92697-3435, USA}

\history{Received on XXXXX; revised on XXXXX; accepted on XXXXX}

\editor{Associate Editor: XXXXXXX}

\maketitle
\begin{abstract}
    In this supplementary material we discuss TAME and its triangle correctness measure as well as give a full chart of the runtimes of all algorithms tested.
\end{abstract}
\section{Triangle Alignment}

Simulated Annealing was built on the assumption of almost every move increasing or decreasing the net energy (score) of a system, just as moving particles in annealing metal increases or decreases the net energy of the metal. In the case of network alignment, this assumes that (almost) every change to the alignment has {\em some} measurable effect on the objective function; there should be some kind of guidance for a large fraction of moves. Some solutions spaces are not like that, but are instead extremely sparse, with many scores of exactly zero interspersed with almost delta-function-like jumps in the score. In such a solution space, almost every move has no effect at all, such as moving from a zero-scoring alignment to another alignment also scoring zero. Even the ``perfect'' objective function, which scores some ``perfect'' alignment with 1 and every other alignment with 0, is impossible to find with random search because each move has no affect on the score; there is no guidance for ``good'' and ``bad'' moves. In such cases, there is little hope of ``converging'' on a good solution since even if one finds oneself temporarily in the vicinity of a delta-function increase in the score, the very nature of the random search (with a non-zero temperature) means we are likely to wander back into zero-score ``flatlands,'' and spend most of our time there. We mentioned these restrictions on simulated annealing in section 2.1.1 of the main paper when introducing SANA.

With the method of moves we have chosen for SANA, one such flatland is ``triangle alignment,'' where the {\em only} thing the objective function cares about is the number of aligned triangles. This is the objective function used in TAME \citep{tame}. With the system of moves we currently use in SANA, we were unable to get a good triangle alignment, for the above reasons. In this sense there can indeed be objectives that SANA may not excel at compared to a hand-coded deterministic algorithm.  We hypothesize that SANA {\em could} easily be modified to do well at aligning triangles.  First, we would exhaustively list all the triangles in each graph, for which asymptotically optimal algorithms exist \citep{Goodrich}; then, we would program SANA's ``moves'' to swap or move entire triangles.  However, we believe this would be a waste of time because the number of triangles in all the biological networks we've encountered involve only a small fraction of all the edges in the networks; we find it hard to believe that {\em any} good alignment algorithm could possibly recover relevant biology by completely ignoring most of the edges in the network.  In the BioGRID networks tested in this paper, the amount of edges participating in triangles ranged from as low as 18\% in the case of CElegans to as high as 92\% in the case of SCerevisiae. Most of the networks have less than half of their edges in triangles, indicating that a {\em majority} of topological information is discarded when using Triangle Conservation. %We will test this hypothesis explicitly in our next paper when we measure, in many ways, the biological value of the alignments produced by various objective functions.  (And in the case of TAME we will use TAME itself to build the alignments.)

\begin{figure}
    \centering
    \includegraphics[width=.5\textwidth]{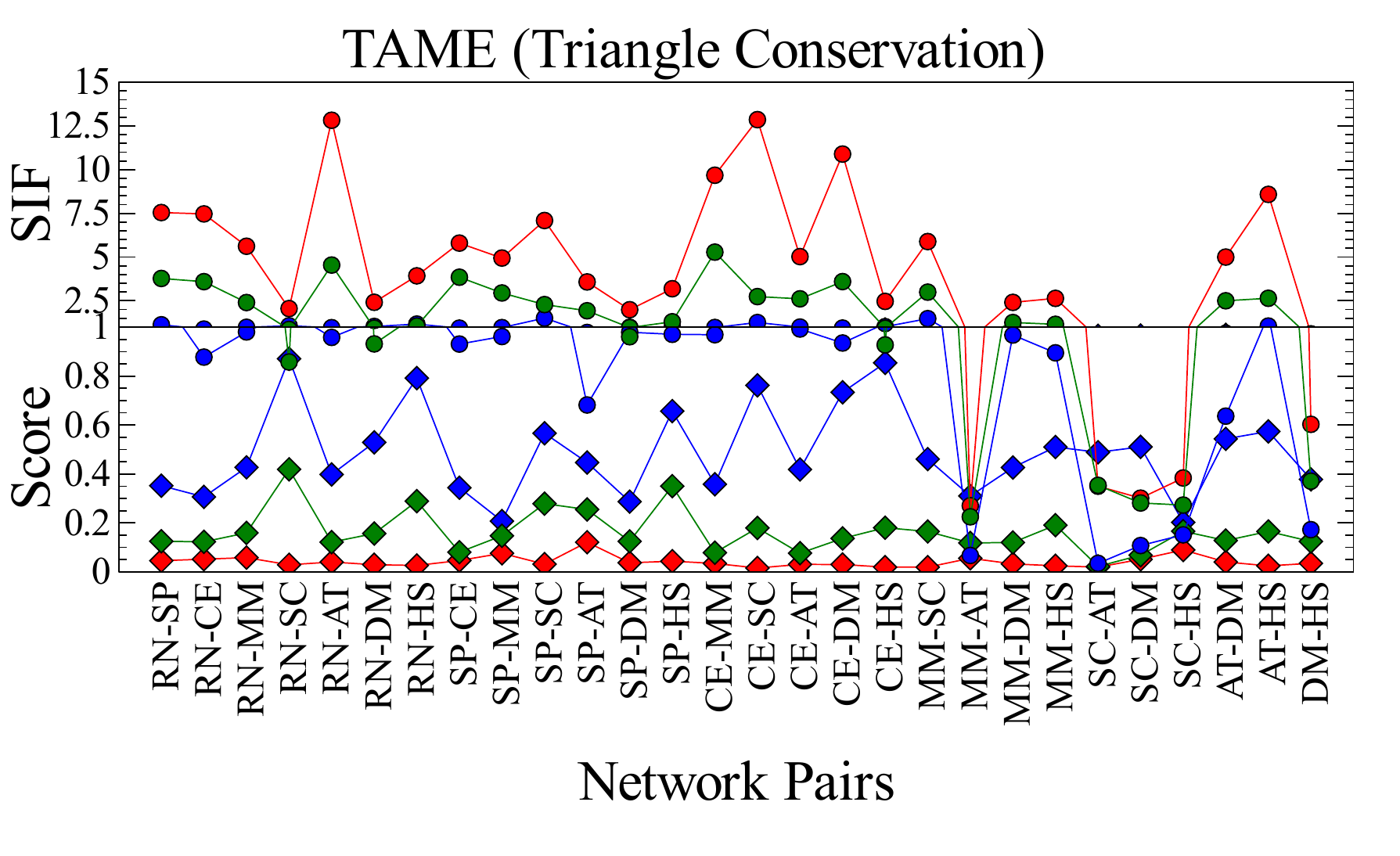}
    \caption{The chart of TAME's comparison against SANA. It follows the same pattern as the first two columns of Figure 1 of the main publication.}
    \label{fig:tamegraphs}
\end{figure}

Figure \ref{fig:tamegraphs} includes the chart for SANA's comparison against only TAME. It follows the key in Figure 1 of the main publication. The chart shown has similar graphical issues around $y = 1$ to NATALIE 2.0 due to the scale for the top half of the chart being very different than the bottom half of the chart.

\section{CPU Analysis}

The network alignment algorithms take a variable amount of time to align networks. Table \ref{tab:cpu} shows how long each network aligner took for each pair of networks. We performed all tests on AMD Opteron 6378 processors. We only count the time that each algorithm took to make the alignment; any overhead or preparation is not included. ModuleAlign is not included on this table because their code is too difficult to run in parallel properly and takes far too long to run in series than is worth it. We originally ran ModuleAlign in series but forgot to measure the exact time and rerunning would take too long than is important. ModuleAlign took around an hour to sometimes more than ten hours.

\begin{table*}
\caption{A complete runtime summary of all comparisons. The network aligners are organized from left to right by speed. All numbers are in seconds. Some network aligners, OptNetAlign and LGRAAL, used a user-specified amount of time.}
\label{tab:cpu}
\centering
\footnotesize
\begin{tabular}{| c c c c c c c c c c c |} 
 \hline
 
 Pair & PROPER & Hubalign & WAVE & SANA$^1$ & GHOST$^2$ & LGRAAL & OptNetAlign & NATALIE 2.0 & GREAT & MAGNA++ \\
\hline
RN-SP & 4.33 & 10.563 & 46.439 & 1200.982 & 218.272 & 3626.067 & 21627.105 & 36318.183 & 1782 & 6588.55 \\
\hline
RN-CE & 2.19 & 14.884 & 67.499 & 1205.878 & 254.168 & 3602.511 & 21594.192 & 36798.131 & 1710.79 & 7946.815 \\
\hline
RN-MM & 7.36 & 18.619 & 89.104 & 1204.306 & 357.098 & 3604.936 & 21597.395 & 37682.798 & 1584.07 & 10036.277 \\
\hline
RN-SC & 141.93 & 86.6 & 108.63 & 1207.698 & 697.682 & 3678.167 & 21620.571 &  &  & 37620.052 \\
\hline
RN-AT & 6.8 & 22.609 & 116.183 & 1202.232 & 494.11 & 3653.807 & 21598.76 & 39227.251 & 2464.4 & 12815.721 \\
\hline
RN-DM & 5.21 & 43.941 & 151.403 & 1202.941 & 761.318 & 3646.642 & 21594.732 & 6252.963 & 4954.68 & 23582.27 \\
\hline
RN-HS & 43.25 & 57.022 &  & 1204.202 & 1576.823 & 3718.981 & 21596.822 &  &  & 59503.783 \\
\hline
SP-CE & 3.08 & 18.746 & 86.451 & 1205.717 & 417.435 & 3618.937 & 21612.816 & 37658.221 & 5909.08 & 9233.363 \\
\hline
SP-MM & 4.35 & 23.241 & 118.53 & 1207.585 & 568.282 & 3621.64 & 21590.237 & 39494.238 & 5926.76 & 11278.529 \\
\hline
SP-SC & 32.42 & 54.946 & 142.788 & 1211.021 & 921.838 & 4035.292 & 21616.769 &  &  & 41411.814 \\
\hline
SP-AT & 5.71 & 32.811 & 147.993 & 1207.724 & 733.266 & 3790.599 & 21632.481 & 9208.076 & 7320.69 & 14183.744 \\
\hline
SP-DM & 7.8 & 57.022 & 206.362 & 1203.379 & 908.696 & 3757.407 & 21650.206 & 12863.58 & 11967.41 & 25254.116 \\
\hline
SP-HS & 31.28 & 95.469 &  & 1204.168 & 2020.892 & 3820.667 & 21644.772 &  &  & 61487.093 \\
\hline
CE-MM & 6.94 & 96.675 & 286.315 & 1208.865 & 934.614 & 3663.708 & 21641.083 & 5541.833 & 15797.56 & 12687.727 \\
\hline
CE-SC & 36.81 & 133.092 & 346.997 & 1211.463 & 1406.377 & 3647.845 & 21619.654 &  &  & 41783.25 \\
\hline
CE-AT & 7.46 & 96.456 & 370.909 & 1203.619 & 1203.285 & 3618.605 & 21620.149 & 6869.122 & 17449.32 & 15844.895 \\
\hline
CE-DM & 16.16 & 139.15 & 481.428 & 1201.269 & 1544.476 & 3720.279 & 21656.884 & 15684.91 & 19398.99 & 26565.248 \\
\hline
CE-HS & 49.9 & 235.182 &  & 1207.018 & 2979.279 & 3750.567 & 21581.82 &  &  & 62952.649 \\
\hline
MM-SC & 247.92 & 290.84 & 667.415 & 1203.642 & 2383.905 & 3890.048 & 21648.963 &  &  & 44150.335 \\
\hline
MM-AT & 22.02 & 269.954 & 770.317 & 1202.983 & 1888.38 & 3902.632 & 21632.147 & 10884.46 & 59376.92 & 18204.402 \\
\hline
MM-DM & 24.14 & 438.859 & 914.006 & 1206.956 & 2518.58 & 3687.144 & 21597.686 & 23804.774 & 66183.56 & 29844.522 \\
\hline
MM-HS & 110.46 & 621.394 &  & 1206.642 & 4655.939 & 5872.466 & 21618.983 &  &  & 65119.34 \\
\hline
SC-AT & 221.57 & 660.456 & 1131.38 & 1211.012 & 7490.71 & 4020.427 & 21760.065 &  &  & 48313.407 \\
\hline
SC-DM & 59.23 & 593.628 & 1495.722 & 1214.728 & 9495.281 & 3799.839 & 21689.732 &  &  & 62692.563 \\
\hline
SC-HS & 176.36 & 1080.103 &  & 1210.358 & 13686.569 & 5248.067 & 21773.541 &  &  & 100999.695 \\
\hline
AT-DM & 46.91 & 648.73 & 1642.546 & 1205.803 & 3653.613 & 3962.908 & 21602.428 & 142527.574 & 239526.32 & 33663.098 \\
\hline
AT-HS & 182.43 & 862.562 &  & 1216.563 & 6687.567 & 4718.554 & 21608.316 &  &  & 68265.577 \\
\hline
DM-HS & 96.58 & 1735.286 &  & 1223.887 & 12888.451 & 6127.944 & 21641.357 &  &  & 82430.435 \\
\hline
Average & 57.164 & 301.387 & 447.067 & 1207.237 & 2976.675 & 3993.096 & 21631.060 & 30721.074 & 30756.837 & 36944.974 \\
\hline
\end{tabular}
$^1$SANA takes a few extra minutes at the start of each run to calculate the temperature schedule. If this is included, SANA takes around 1400 seconds.

$^2$GHOST took almost 2 years of CPU time (11 days on 64 cores) in some cases to make the spectral signatures for individual networks. This time is not included in this chart.
\end{table*}

%I dont think it's worth it to put the chart (split_supplement.pdf) here because it doesn't serve a real purpose other than to clearly outline that they beat us, which isn't good. It's bad PR

\bibliography{document}{}
\bibliographystyle{natbib}